Author(s): Nurhas, Irawan; Jahanbin, Pouyan; Pawlowski, Jan; Wingreen, Stephen; Geisler, Stefan

Title: Patterns of Sociotechnical Design Preferences of Chatbots for Intergenerational Collaborative Innovation : A Q Methodology Study

Year: 2022

Version: Published version

Copyright: © 2022 Irawan Nurhas et al.

Rights: CC BY 4.0

Rights url: https://creativecommons.org/licenses/by/4.0/

Please cite the original version:

Nurhas, I., Jahanbin, P., Pawlowski, J., Wingreen, S., & Geisler, S. (2022). Patterns of Sociotechnical Design Preferences of Chatbots for Intergenerational Collaborative Innovation : A Q Methodology Study. Human behavior and emerging technologies, 2022, Article 8206503. https://doi.org/10.1155/2022/8206503

Research Article

# Patterns of Sociotechnical Design Preferences of Chatbots for Intergenerational Collaborative Innovation: A Q Methodology Study

**Irawan Nurhas** [1,2] **Pouyan Jahanbin,**[3] **Jan Pawlowski** [1,2] **Stephen Wingreen** [3] **and Stefan Geisler** [1]

[1]*Institute of Positive Computing, Ruhr West University of Applied Sciences, Bottrop, Germany*
[2]*Faculty of Information Technology, University of Jyväskylä, Jyväskylä, Finland*
[3]*Department of Accounting and Information Systems, University of Canterbury, Christchurch, New Zealand*

Correspondence should be addressed to Irawan Nurhas; irawan.nurhas@hs-ruhrwest.de

Chatbot technology is increasingly emerging as a virtual assistant. Chatbots could allow individuals and organizations to accomplish objectives that are currently not fully optimized for collaboration across an intergenerational context. This paper explores the preferences of chatbots as a companion in intergenerational innovation. The Q methodology was used to investigate different types of collaborators and determine how different choices occur between collaborators that merge the problem and solution domains of chatbots' design within intergenerational settings. The study's findings reveal that various chatbot design priorities are more diverse among younger adults than senior adults. Additionally, our research further outlines the principles of chatbot design and how chatbots will support both generations. This research is the first step towards cultivating a deeper understanding of different age groups' subjective design preferences for chatbots functioning as a companion in the workplace. Moreover, this study demonstrates how the Q methodology can guide technological development by shifting the approach from an age-focused design to a common goal-oriented design within a multigenerational context.

## 1. Introduction

Intergenerational collaboration facilitates the transfer of knowledge in the pursuit of innovation. Intergenerational collaboration is a type of social interaction between people of different ages working together for a specific goal. People of different ages may be classified within a particular context, such as a family or corporate context [1–3], for different reasons, such as wars, economic recessions [4, 5], or technological backgrounds—digital natives vs. digital immigrants [6, 7]. People of different ages can have different levels of knowledge, expertise, and, above all, the experience that they can share for the mutual benefit of collaboration. Through social inclusion in the workplace, the intergenerational collaboration will lead to innovative practices and sustainable enterprises as well as competitive business models in the digital age [1, 8, 9]. Moreover, intergenerational collaboration often benefits all generations in terms of well-being and the growth of human potential [10–13]. Despite the various technological interventions that have facilitated collaboration between individuals from different generations [14–16], there are still several challenges and a digital divide that may hinder technology-based intergenerational collaboration [9, 11, 17]. Numerous impediments must be addressed, including ambiguity in communication [18], different cultural and technological backgrounds that make managing virtual activities and cocreation difficult, or determining the best possible time for all parties to (virtually) meet [19, 20]. However, the study shows that chatbots can assist with such technical and managerial tasks [20]. Likewise, in creative activities such as ideation and cocreation, where divergent perceptions or ways of thinking can hinder creativity, chatbots can assist through social facilitation or inhibition [21] and system customization [22, 23].

In this study, we investigate the potential of chatbots to facilitate intergenerational collaboration: more specifically, collaboration in innovative activities such as ideation, design, and development. A chatbot is a social robot-, voice-, or text-based conversation system widely used to help users adopt certain services or technologies [15, 24]. The chatbot can also enhance virtual learning experiences [25, 26], as a creative teammate in the workplace [27–29]. Due to the exponential development of research in artificial intelligence (AI), chatbot technology is also improving. Over the last decade, chatbots have been integrated into smartphones as a personal assistant that provides suggestions and relevant information [15, 24]. Because the "age" factor is subjectively perceived in the workplace and influences the control of attributes and motivation [30], personal emotional experiences and subjective preferences have a significant impact on the use of enabling technologies (such as chatbots as virtual machines for social interaction [3]) for intergenerational collaboration [31–34], especially among older adults [3, 35]. Therefore, chatbots should be developed by identifying interaction patterns and involving users in cocreating products and services. This can increase the subjectivity of products and services, allowing for better customization [22].

Notwithstanding the advantages of chatbots that offer human-like competencies [26, 36, 37] and provide human-like responses [26, 37–39], studies of chatbots focus primarily on technological aspects [15, 25] and seem to lack an understanding of group-subjective design preferences [40], especially in an intergenerational context. Therefore, it is crucial to understand human preferences as a fundamental aspect of design technologies for improved human-system interactions [41] with chatbots [38]. In this study, we explored the structural elements of chatbot design in an intergenerational context by addressing the following research questions: What types of intergenerational collaborators emerge as chatbots support digital workplace innovation processes? How should chatbots be developed to support collaboration across generations?

We conducted a Q methodology analysis as an impression method in order to address the research questions and develop an understanding of the various types of collaborators based on chatbots' subjective preferences [42, 43]. The Q methodology systematically maps cognitive experiences onto a topic [44, 45] or pattern with a similar perspective or subjectivity in a person or group of people [42, 43, 46].

By following the logic of abductive reasoning as the critical approach to knowledge discovery [31], this study marks the first step towards gaining a deep understanding of chatbot design based on the systematic observation of subjective preferences in an intergenerational context and among four types of chatbot user groups. In this sense, the study contributes significantly by broadening our understanding of a series of chatbots' specialized design concepts and addressing their technological and social dimensions. Furthermore, each collaborator's typology consists of diverse goals and priorities that we argue can be used in a problem-based design or lean software development strategy [41, 47–49].

## 2. Theoretical Background

In this study, intergenerational innovation is understood as a shared work-related endeavor between younger and senior adults with an age difference of around 20 years [11, 50]. The disparity in years may be a potential source of technological discrepancies that hinder collaboration [50]. However, generational diversity strengthens organizational diversity and provides opportunities to develop collaborative innovation and learning sources [8, 9, 11]. Since innovation is a prerequisite for a successful organization, studies have empirically proven how generational diversity promotes innovation and creates a decade-long sustainable business model [1, 8, 9].

Generational disparities concern the difference in experience [51] and the social consequences of a bipolar split [52] more, avoiding psychological and physical stigma [53] as well as the influence of subjective age on motivation at work and the control of work-related attributes [30]. Therefore, diverse user-subjective viewpoints, preferences, goals, and motivations are necessary to a create reliable group or user-system interactions based on goal-oriented system design [14, 40, 41]. Using technologies to foster intergenerational innovation has met socio-technological hurdles in cultural and institutional domains [11, 14, 19, 51]. Recognizing and addressing these problems [47] are necessary to promote social participation in technical design and the overall technology design process [33, 52, 53].

Moreover, in the early stages of product or service creation, the importance of defining design preferences for meaningful objectives is highly subjective [54]. To determine priorities, the identification of design preferences should also be thorough and systematic [43, 54, 55]. A positive approach to incorporating well-being and sociotechnical interventions in which technology can be generated inevitably leads to questions about awareness of subjective preferences [56], including the design of social and technical chatbots with unique barriers or challenges. On the one hand, previous studies have established multidimensional barriers to digital collaboration [11, 19, 57, 58]. The barrier dimensions focus on the context of the collaborative innovation process, where creativity and the exchange of ideas between different age groups for products or services design are at the heart of the collaboration [19].

The first barrier dimension is the perceptual barrier, which refers to the obstacles in intergenerational collaboration due to differences in perceptions and negative perspectives toward other generations. Perceptual differences can arise from differences in experience, domain knowledge background, and mindset and from ignorance of intergenerational differences or inability to collaborate with different ages [19, 51, 58].

The second barrier is technical and operational. Here, obstacles arise from differences in familiarity with and operation of various tools and technologies used in collaboration. The complexity of the shared technology may result in inadequate access to the technology or a lack of training in using the technology [17, 19, 59].

Emotional barriers form the third dimension. Unlike the first dimension, which focuses on how others see one another, the third dimension is about self-perception: how one sees oneself as a member of a particular generation interacting with others from other generations. This barrier can arise from differences in physical or cognitive functioning, fear of the technology used in collaboration, feelings of isolation, lack of appreciation, or underestimation by previous generations. Even if particular generations or collaborators are familiar with the technology, they may lack confidence in their (technical) ability to collaborate with other generations [16, 19, 59].

The fourth dimension, barrier, is the dimension of obstacles and barriers that arise from cultural differences between the collaborating generations. These cultural differences are external factors related to unwritten rules and things like feminism, long- vs. short-term orientation, and power distance that affects collaboration [60]. The cultural barrier dimension includes the lack of awareness for cultural differences in intergenerational collaboration, and the development of technologies does not take into account cultural aspects such as the use of certain symbols and colors or a distinct understanding of the traditional culture of one generation compared to other generations [19, 57, 58]. Different cultural barriers can hinder collaboration, but they can also serve as a source of innovation and enrich the exchange of information between employees.

The last factor is the dimension of institutional barriers. This dimension is also an external barrier, but it is administrative and tied to organizational norms often documented. This barrier complements the cultural dimension by focusing on the underlying issues of the organization in which the collaboration takes place. Some of the institutional barriers that can occur are privacy policies, different educational backgrounds, disparate team locations, and how resources are distributed within the organization, whether closed or open. These various factors can combine to form a layer of institutional barriers that impede intergenerational collaboration in the innovation process [3, 19, 61].

On the other hand, implementing system design interventions that positively impact user experiences often relies on technological choices [41, 52, 56] within an intergenerational context. For example, previous research has considered using chatbots as the technology to promote social inclusion [38]. Furthermore, significant improvements of AI push the boundaries of chatbot design limits, allowing for more natural interactions and better visual and audio representations [36, 62, 63]. This ongoing evolution of chatbots creates more opportunities for their use as companions in various settings, including a potential technology for intergenerational collaboration [3, 64]. The various functions and applications of chatbots that can support intergenerational innovation processes include the following: (1) reducing communicative ambiguity in workplace contexts [18], which can aid in the removal of perceptual barriers by restricting broad topic interpretation and selection options, (2) supporting problem-based learning individualization [23] and collaborative distance learning [65], which can help overcome emotional, cultural, and institutional barriers to learning more about a topic together, and through artificial intelligence and machine learning in a more personal setting [22, 23], focusing on identifying common problems and raising awareness of intergenerational differences, and (3) facilitation of debate and consensus building [65], presentation of documents and results, management and administrative support in the cocreation process, and the ability to guide group collaboration in the context of professional work through the aesthetic design and various group work features [66]. These features and functions can help solve technical and operational challenges and promote and support critical issues such as privacy and information credential in the context of institutional barriers.

Inequalities in the use of chatbots between generations are also increasingly eradicated from a social inclusion standpoint. This is accomplished through the development of chatbots with the goal of bridging the digital divide, particularly for older adults [67, 68]. Through intangible interactions such as speech and physical interactions such as portable physical devices and the use of light colors, chatbots have been designed to provide a sense of companionship and well-being to older people [67]. Subsequently, the integration of facial recognition and sentiment analysis enhances older adults' ability to extract and conceptualize technical requirements to minimize the complexity of technological engagement. These two approaches result in a gamified chatbot capable of empathetically responding to the user's emotional identification and presence during the automated engagement [68]. With the human-machine interaction research area continuing to develop, chatbots also have the opportunity to emulate natural communication and networking capabilities [26, 36, 37]. As such, one study by Skjuve and Brandzaeg [37] suggests ten interactive chatbot communication capabilities used as a framework for this study:

(i) Chatbot self-disclosure is a chatbot's ability to share knowledge, emotions, or experiences

(ii) Chatbot empathy is a chatbot's capacity to grasp receiver emotions

(iii) Chatbot social relaxation is a chatbot's ability to provide user-friendly contact

(iv) Chatbot interaction management is a chatbot's ability to handle contact by negotiating the topic to be addressed, understanding nonverbal user emoji messages, or effortlessly managing changes in conversation topics

(v) Chatbot assertiveness is a chatbot's ability to stand unilaterally for its rights without violating users' rights

(vi) Chatbot altercentrism is a chatbot's ability to provide social security using relevant questions

(vii) Chatbot expressiveness is a chatbot's ability to express verbal or nonverbal feelings

(viii) Chatbot supportiveness is a chatbot's ability to defer unfavorable opinions and assist user-equal emotions

(ix) Chatbot immediateness is a chatbot's ability to open communication

(x) Chatbot environmental control is a chatbot's ability to achieve tasks and aims by precisely defining its capacities and limitations

Moreover, the aim of designing technology for workforce diversity is at accelerating the process of social inclusion [52] by integrating the determinants of well-being into the technology's design [11, 56]. Previous research has summarized the various factors that promote social participation in intergenerational collaborative technology development regarding the social dimension in which technology may contribute to well-being determinants. For instance, a collaboration fosters pride and common values, sharing ideas through stimulating events, and lastly, an explorative, problem-oriented, and collaborative atmosphere for learning [1, 2, 11, 14, 16, 51, 59, 69]. We will now present this study's research methodology to understand how the generational preferences correlate and create a different user form for chatbots as companions in intergenerational contexts.

## 3. Methodology

In this study, the Q methodology [42] was used to uncover chatbots' preferences for facilitating intergenerational collaborative innovation systematically. The Q methodology supports exploratory and mixed-method analysis [46, 70, 71], which suits our study's aim to understand chatbots' design preferences for individuals and groups. The essence of Q methodology studies is ideally defined by questions about the phenomenon under discussion, which need to be answered before a more general theory can be established [43, 55, 72]. Because of this, the Q methodology naturally lends itself to the use of abductive reasoning [55, 72]. This approach encourages the development of hypotheses based on factual observations of system preferences [43, 45, 55] concerning chatbot use in an intergenerational workspace.

Stephenson Stephenson [42] developed the Q methodology to gain scientific insight into subjective viewpoints through Q sorting, Q factor correlation, and Q factor analysis [42, 43, 55]. The approach empirically clusters participants' Q sorts based on their pattern of preferences [43, 55, 71]. The Q methodology begins by developing a set of statements (Q set) guided by the theory of concourse, or the theory of communicability constructs a set of concepts reflecting preferences [42, 43, 55]. The Q set can be collected from direct interviews, expert opinions, websites, social media, or literature [45, 55]. Figure 1 visually depicts the process of the Q methodology, from concourse development to the Q set, Q sorting, and factor analysis.

It is important to note that in Q methodology research, the concourse is the subject of the study rather than a population of people. Therefore, the concourse is sampled as described (the Q set) and measured by a set of people (see the P set) using the Q sort procedure. Ideally, the P sets are chosen for their ability to instantiate the subjective perspectives of interest to the research, in our case, the intergenerational differences of chatbot design preferences. The factor analysis is then performed on the Q sorts to reveal the structure of the concourse [55, 71].

*3.1. Listing the Q Set.* The list of statements (i.e., Q set) has been developed based on a literature review, feedback from people of interest (members of the younger and older generations), and discussions among the researchers. Chatbot competencies applicable to chatbot design principles [37] have been adopted to guide Q set development, to construct a Q set that is a representative sample of the concourse of chatbot design. The proposed principles focus on the communication skills necessary for human-system interactions and facilitating digital group collaboration [37, 39, 73]. Intergenerational barriers and enablers [1, 2, 10, 11, 14, 16, 19, 51, 59, 69] were used as a framework for activities either promoting or hindering intergenerational collaboration. All concepts from the chatbot communication capability framework were attempted to be represented in the Q set [43, 55], as shown in Table 1. The Q set was used to establish preference representations of the study context that can capture motivation [19, 41, 74], including (personal) gains and pain factors, for collaboration in intergenerational settings. The Q set presented in Table 1 represents a problem domain (with Id: s11 to s15) and a solution domain, divided into a technical approach (Id: s1 to s10) and an approach to social interaction, respectively (Id: s16 to s19).

*3.2. Study Participants (P Set).* We pursued the concept of intergenerational collaboration with an age difference of around 20 years [11, 50]. Therefore, we divided the P set into senior and younger adults. Eleven participants were selected to represent younger adults from academics and students involved (coded with Id P1 to P11, age representation 18–23 years: 45.45%, 23–28: 18.18%, 28–33: 18.18%, 37–41: 18.18%). For the senior adults, thirteen participants participated in this study with age above 55 years, who still regularly use the Internet and online media. The senior adult participants were coded with Id SP1 to SP15 and age representation (55–60: 69.23%, 60–65: 23.08%, >65%: 7.69%). Since the focus of the study was preferences between different senior and younger adults and the proposal of a chatbot to support intergenerational collaboration, for this study, only age differences between senior and younger adults were considered. However, to put the P set in the same context, a descriptive scenario called a "condition of instruction" was highlighted and presented to the P set [43, 55, 71], in which members of the P set were asked about their preferences if a chatbot was used to facilitate digital collaboration with other generations.

Based on the concourse theory, the goal is not to obtain large numbers of participants to describe the study's population but to get a representative sample of Q statements representing the concourse [43, 55]. Therefore, the Q method does not require many participants to obtain a good result, although careful development of the Q set is important. A successful outcome can be obtained with fewer (even less than 30) participants [43, 44, 55]. The goal is to make generalizations about the concourse rather than a population of people. Still, researchers can also make generalizations with other more common survey methods if the goals of the research require it [75].

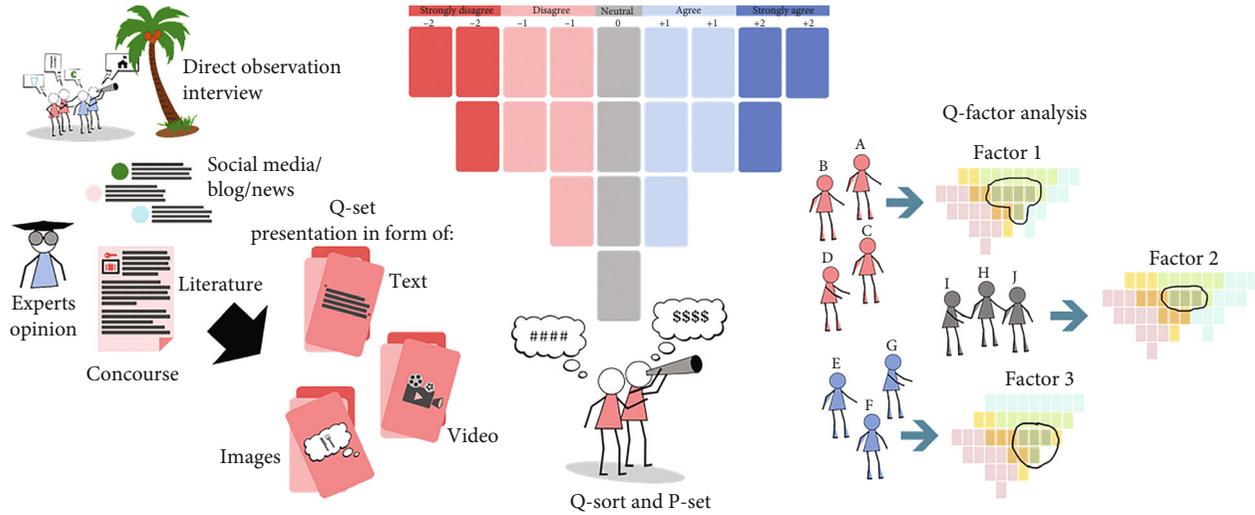

Figure 1: An illustration of the process of the Q methodology, from the concourse development to factor analysis.

3.3. Q Sorting and Factor Analysis. The participants (P set) of the Q sorting procedure were asked to sort the statements by comparing each statement with other statements in the Q set [43, 71]. An online Q sort using htmlQ (https://github.com/aproxima/htmlq) and firebase (https://firebase.google.com/) was applied to centralize the collected data and provide flexibility for the participants. Following the Q methodology guidelines [43, 55, 71], the P set first differentiated the statements into three piles (see Figure 2(a)) to easily compare each statement with other statements before placing the former into the Q sort distribution, which resembles a near-normal distribution consisting of 1 place for both an extreme positive ("much closer to what I think") and negative ("less close to what I think"), two places for −2 and +2, four places for statement categories −1 and +1, and five places for neutral (see Figure 2(b)). After sorting each statement, the participants were given a chance to review the placement of the Q set. We've added open-ended questions to help us in determining the reason behind the statement's order for the extreme negative and positive [43, 55]. Finally, we asked members of the P set to provide information about their age and years of experience in collaborating in an intergenerational setting as filtering questions. 24 out of 25 responses(Q sorts) we received met our filtering criteria regarding experience working with other generations.

We performed the statistical analysis with a specifically designed software package called PQMethod [55, 76], which allows researchers to choose extraction and rotation techniques for their Q studies' factor analysis [77]. Rotations are performed using either objective statistical tools or through the employment of subjective and conceptual deliberations or a specific theory of interest to the researcher(s) [78]. Researchers in Q methodological studies usually use the centroid factor analysis (CFA) and hand rotation, sometimes called "judgmental rotation" because of its reliance on the researcher's judgment to manually rotate the factor axes to obtain an interesting and interpretable solution. This approach is driven by the research questions and goals, albeit subjectively, allowing researchers to investigate and analyze data relying on theoretical and conceptual deliberations [79]. However, principal component analysis (PCA) employment and varimax rotation have appeared in many Q studies [80]. PQMethod allows the incorporation of both statistical rules and theoretical, conceptual interpretations to support data analysis.

In our study, using PQMethod, a correlation matrix was initially generated by measuring the extent of the interrelationship between each Q sort with every other Q sort [55]. A factor analysis was performed using the correlation matrix in the following step. Since the CFA and PCA are both commonly employed for implementing factor analysis in Q methodological studies, we tested both methods [55] in parallel, because both methods have their advantages and disadvantages, with the intent of proceeding with the solution which provided the most insight into our research questions.

The PCA generated eight factors from the correlation matrix using the Kaiser-Guttman criterion, which retains seven factors with eigenvalues greater than 1 [55]. According to Brown [79], a factor is retained if it contains at least two significant factor loadings following extraction. At the 0.05 level of significance, with loadings exceeding 1.96, the standard error is significant by the following criteria:

$$\lambda = 1.96 \times \frac{1}{\sqrt{\text{no. of items in Q set}}} = 1.96 \times \frac{1}{\sqrt{19}} = 0.45. \quad (1)$$

Six factors were extracted using this standard. Humphrey's rule, which recommends retaining a factor only if the crossproduct of its two highest loadings (ignoring sign)

Table 1: List of statements related to the design of chatbots for intergenerational innovation.

| Id | Concept | Example of statement | References |
|---|---|---|---|
| | | Technological (chatbot) interventions | |
| s1 | Self-disclosure | The chatbot should be able to share personal thoughts or experiences. | [37] |
| s2 | Empathy | The chatbot should demonstrate that it understands and sympathizes with the user when appropriate. | [37] |
| s3 | Social relaxation | The chatbot should feel comfortable and secure during the interaction and not be anxious. | [37] |
| s4 | Interaction management | The chatbot should demonstrate turn taking and discuss and develop different topics. | [37] |
| s5 | Assertiveness | The chatbot should be able to stand up for itself and its rights but at the same time be accustomed to and not violate the user's rights. | [37] |
| s6 | Altercentrism | The chatbot should make the user feel that it is interested in what they have to say, ask appropriate questions, be polite, and display appropriate emotional expressions and content. | [37] |
| s7 | Expressiveness | The chatbot should be able to express its feelings either verbally (e.g., laughter) or nonverbally (i.e., through emojis). | [37] |
| s8 | Immediacy | The chatbot should be available and open for communication. | [37] |
| s9 | Supportiveness | The chatbot should not judge the user and make the user feel equal. | [37] |
| s10 | Environmental control | The chatbot should be able to accomplish its goals and objectives. | [37] |
| | | Barriers that can hinder intergenerational collaboration | |
| s11 | Negative judgment towards other | Digital collaboration with different generations will be difficult due to other generations' limitations (cognitive or physical). | [11, 51, 58] |
| s12 | Technical/operational differences | I think that digital collaboration will be difficult due to different technological experiences. | [11, 17, 59] |
| s13 | Emotional barriers | I feel emotionally unsafe collaborating digitally with other generations. | [11, 16, 59] |
| s14 | Cultural differences | Cultural differences hinder digital collaboration between generations. | [11, 57, 58] |
| s15 | Institutional challenges | Due to administrative barriers (i.e., different organizational rules and requirements), collaboration with other generations will be difficult. | [3, 11, 61] |
| | | Activities that can trigger intergenerational collaboration | |
| s16 | Joy and playfulness | It is important to me that intergenerational digital collaboration offers a playful environment. | [11, 16, 69] |
| s17 | Exploration and interest | It is important that I can work with other generations in a problem-oriented learning environment. | [11, 69] |
| s18 | Achievement of collective goals | I think it is important that I can work with other generations based on common goals. | [11, 59] |
| s19 | Apprenticeship | I think it is important that I work together with other generations through apprenticeships. | [2, 11] |

exceeds twice the standard error [79], is calculated as follows:

$$\text{Standard error} = \frac{1}{\sqrt{\text{no. of items in Q set}}} = \frac{1}{\sqrt{19}} = 0.23. \quad (2)$$

Twice, the standard error for our study is 0.46 (2 × 0.23 = 0.46). Applying this rule, only two of our factors met this criterion (factors 1 and 3). As such, we, therefore, opted for a less stringent application of the Humphrey rule, which is to insist that the crossproducts exceed at least the standard error [79]. Under these further circumstances, the extraction of five factors would clearly be acceptable since three additional factors (the crossproducts of factors 2, 4, and 5 then exceed the standard error of 0.23) were included along with factors 1 and 3.

Following the factor analysis, we applied a varimax rotation and created a factor array for each factor. A factor array represents the "average" Q sort associated with each factor [55]. In order to gauge the PCA and CFA results, we then repeated the aforementioned procedure for defining the appropriate number of factors using a CFA. Initially, the CFA created seven factors using the correlation matrix, from which we retained six factors with eigenvalues greater than 1; then, following Brown's and Humphrey's rules, the procedure resulted in a 3-factor solution, which could explain considerably less variance compared to the PCA solution. Thus, we proceeded with the PCA solution for further analysis.

Initially, varimax was used mainly to explore patterns of interest in the data. Since the primary study aim was at ascertaining the central or dominant viewpoints within the participant group, a series of hand adjustments followed to ensure that the maximum possible number of participants was included in the factors and that the factors were interpretable. Since the goal of Q method research is to understand subjective viewpoints of people, interpretability of a factor is also used as a criterion for whether a factor is retained or rejected. The fifth factor, although statistically significant, did not represent a pattern of thought distinct from the first four and therefore was removed from further consideration. As shown in Table 2, the correlation between the factors is weak to moderate.

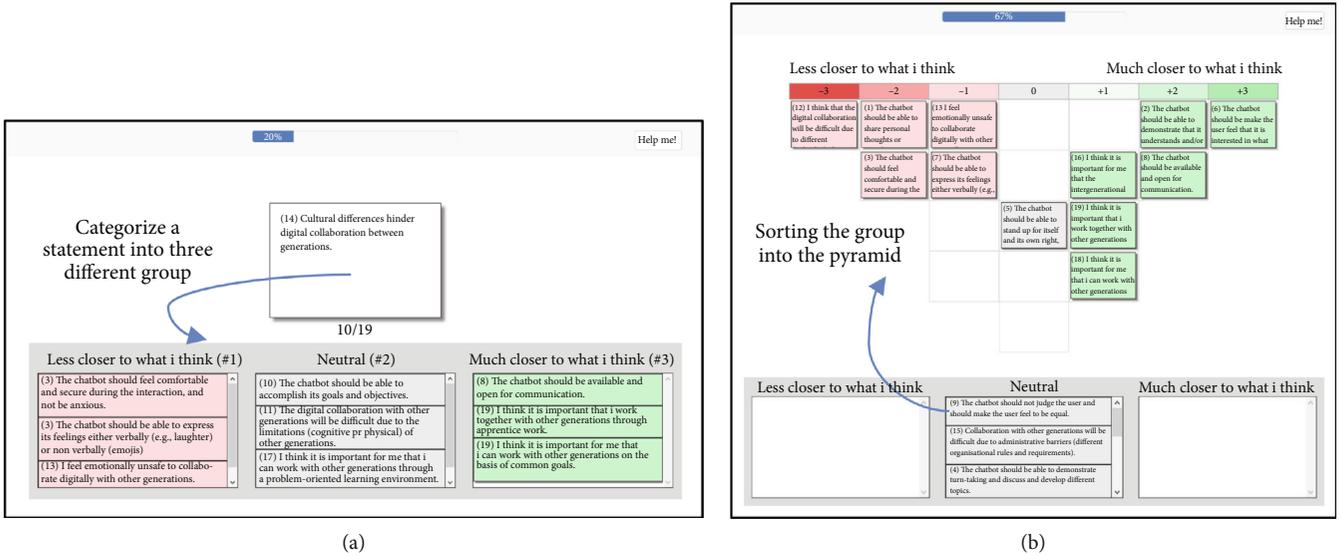

Figure 2: Design of the online Q sorting: (a) presorting by randomly displaying the statement on the white card; then, the user categorized the statement into three different categories; (b) Q sorting pyramid consisting of some sorted statements.

## 4. Results

In this section, we report the results and descriptions of the four factors retained by the above-discussed criteria. Factor descriptions are characterized by the subjective priorities expressed in the "average Q sort" of the factor array, which in our case are the subjective preferences or attitudes towards the design of chatbot technology. Table 3 reports the factor loadings associated with each Q sort [55], which expresses how each Q sort is associated with each factor extracted. The resulting 4-factor solution with no less than three Q sorts per factor explains 65% of the cumulative variance in the data and hosts 24 Q sorts out of 24, which satisfies the minimum requirement of 60% suggested by previous research [81].

Table 4 reports the factor scores for the 4-factor solution, sorted by consensus vs. disagreement. The highest-consensus statements are at the top side of Table 4, while those with the highest disagreement between P sets are located at the bottom of Table 4. The statements with the greatest consensus statements may be regarded as the values shared by most chatbot users across both generations. In contrast, the statements with the greatest disagreement represent competing values that define generational differences. Factor types are defined according to the factor score associated with the most positively and negatively ranked statements.

Table 2: Correlations between factor scores.

| Factor correlation | Factor 1 | Factor 2 | Factor 3 | Factor 4 |
| --- | --- | --- | --- | --- |
| Factor 1 | 1 | 0.4092 | 0.1893 | 0.2228 |
| Factor 2 | 0.4092 | 1 | -0.0094 | 0.2544 |
| Factor 3 | 0.1893 | -0.0094 | 1 | 0.0349 |
| Factor 4 | 0.2228 | 0.2544 | 0.0349 | 1 |

*4.1. Type of Attitudes towards Chatbot for Intergenerational Collaboration.* In the Q methodology, abductive reasoning describes unexpected empirical facts arising from statistical factor analysis. Consequently, the interpretation of factors is the focus of the analysis [43, 55], which must be codified to provide plausible theoretical reasons for a factor's occurrence [55]. Taken together, in intergenerational settings, the four factors or types of chatbot personas are as follows:

*4.1.1. Type 1: Enthusiastic Senior Collaborators.* Type 1 has explained 21 percent of the variance. Type 1 consisted exclusively of ten senior-adult representatives and therefore reflected their viewpoint on the positivism of intergenerational collaboration and the usage of chatbots as facilitators of digital collaboration. Type 1 is very appreciative if chatbots can sufficiently demonstrate emotional cues and ask questions politely. Type 1 is distinctly different from other types, particularly in terms of its support for chatbots by eliminating negative judgments and allowing all generations to feel equal in digital collaborations. The following main statements reflect these priorities:

(i) s6: (altercentrism) chatbot should make the user feel that it is interested…(+3). Commentaries on why participants in type 1 found the statement highly

Table 3: Factor characteristics and correlations.

| Q sort from the P set (participants) | Factor 1 | Factor 2 | Factor 3 | Factor 4 |
| --- | --- | --- | --- | --- |
| P1 | 0.3115 | 0.246 | **0.7028** | −0.2351 |
| P2 | −0.1315 | 0.4404 | **−0.7214** | 0.1096 |
| P3 | 0.0222 | 0.3019 | 0.5303 | **0.6585** |
| P4 | 0.2606 | 0.1781 | −0.202 | **0.7097** |
| P5 | −0.0443 | **0.6349** | −0.2579 | 0.0746 |
| P6 | −0.1348 | 0.1723 | 0.0308 | **0.6194** |
| P7 | −0.0365 | **0.5992** | −0.1914 | 0.23 |
| P8 | −0.1493 | −0.0246 | **0.6792** | 0.1844 |
| P9 | 0.0560 | **0.6621** | 0.0513 | −0.2639 |
| P10 | 0.2398 | **0.6893** | −0.1134 | −0.1844 |
| P11 | 0.2266 | **0.6114** | 0.3242 | 0.1152 |
| SP1 | **0.6222** | 0.3338 | 0.3309 | 0.1984 |
| SP2 | **0.7929** | −0.0172 | −0.1723 | 0.3416 |
| SP3 | **0.8329** | −0.0904 | 0.0159 | −0.2626 |
| SP4 | **0.7509** | 0.4520 | −0.0211 | −0.0047 |
| SP5 | **0.6237** | 0.2533 | 0.4485 | 0.1348 |
| SP6 | 0.0464 | **0.7225** | 0.0874 | −0.4366 |
| SP7 | 0.3173 | **0.5453** | 0.0846 | 0.3502 |
| SP8 | **0.6645** | 0.6111 | 0.0934 | 0.2215 |
| SP9 | **0.6564** | 0.4127 | 0.2911 | 0.2732 |
| SP10 | **0.7100** | 0.4306 | 0.1301 | −0.0747 |
| SP11 | **0.6576** | 0.0468 | 0.0924 | −0.1408 |
| SP12 | 0.2836 | **0.7251** | 0.2210 | 0.1141 |
| SP13 | **0.6755** | −0.3866 | −0.2257 | −0.0197 |
| % expl.Var. | 23% | 21% | 11% | 10% |

The PQMethod software identifies "exemplar" factor loadings if (1) the factor explains more than half of the common variance ($a2 > h2/2$) and (2) the factor loading is significant at $\alpha < 0.05$ ($a > 1.96/\text{SQRT}(n\text{items})$).

positive are as follows: (SP9): *"I think it's important for a person to feel that they are talking to someone who understands what they are trying to say and is polite and helpful."* (SP10): *"if the user does not feel respected, the whole project loses its value and meaning. It is imperative that people are treated with respect; otherwise, there is no point in going ahead with any of this because it will turn into a situation where users feel neglected, prejudiced against, disrespected, mean spirited"*

(ii) s17: (exploration and interest) I think it is important for me that I can work with others…(+2)

(iii) s9: (supportiveness) chatbot should not judge the user and should make us…(+2)

Members of type 1 consider technological obstacles to be the least important. Type 1 also emphasizes cultural barriers (−1) and negative judgments about other generations (−1) relative to other barriers that may impede collaboration. This viewpoint is expressed in a highly negative (or less relevant) prioritization of the following statements:

(i) s12: (technical/operational barriers) I think that digital collaboration will be difficult due to…(−3). Some available comments as to why the participant in type 1 considered the statement extremely negative are as follows: (SP2): *"this never occurred to me. I don't think technology will cause a problem."* (SP3): *"I believe that all ages can adapt to technology and perform well."* (SP4): *"Just because someone is older or younger does not mean they do not have the same tech. experiences"*

(ii) s13: (emotional barriers) I feel emotionally unsafe to collaborate digitally with others…(−2)

(iii) s15: (institutional barriers) collaboration with other generations will be difficult due…(−2)

All type 1 priorities that received a strong negative rating are barriers, indicating that the emphasis is largely on discussing solutions and interventions. Type 1 shows that some statements are also unique for type 1, particularly a collaboration based on exploration and interest in a playful environment. Distinguished statements for type 1 are as follows: S9 (supportiveness: +2), S17 (exploration and interest: +2), S16 (joy and playfulness: 0), S7 (expressiveness: 0), S11 (negative judgments toward others: −1), S15 (institutional challenges: −2), and S12 (technical/operational barriers: −3).

*4.1.2. Type 2: Thoughtful, Compassionate Collaborators.* Type 2 explains 21% of the study's variance and is correlated with five younger adults and three senior adults. The chatbot design for type 2 should promote crossgenerational collaboration efforts based on shared objectives. Furthermore, chatbots can eradicate derogatory judgments about a single generation and support all generations equally in promoting collaborative practices. The chatbot should also communicate its capabilities and weaknesses. Type 2's top-ranking statements are as follows:

(i) s9: (supportiveness) chatbot should not judge the user and should make…(+3). Some available comments as to why the participant in type 2 considered the statement extremely positive are as follows: (SP6): *"A chatbot is a facilitator, not a judge."* (SP7): *"Because I think sometimes the Ai will infer that it is more intelligent, leaving the user feeling incompetent"*

(ii) s18: (achievement of collective goals) I think it is important for me that I can work with others…(+2)

(iii) s10: (environmental control) chatbot should be able to accomplish its goals and object…(+2)

Compared to type 1, which has three different barriers which scored the highest for lowest priority, type 2 only has one barrier, the emotional barrier, which is among the highest-ranked low-priority statements. Other barriers are in neutral positions in the Q sort distribution. Type 2 takes intergenerational collaboration more seriously than type 1 by sorting the fun and play statements with negative priorities. Chatbots' ability to share actual intergenerational collaboration opinions is also less important to type 2 than other design interventions. Type 2's lower-ranked statements are as follows:

(i) s1: (self-disclosure) chatbot should be able to share personal thoughts or…(−3). Comments as to why the participant in type 2 considered the statement extremely negative are as follows: (P11): *"The chatbot should not have any thoughts of his own."* (SP12): *"I really don't think the chatbot is capable of having personal thoughts or feelings"*

(ii) s16: (joy and playfulness) I think it is important for me that the intergenerational…(−2)

(iii) s13: (emotional barriers) I feel emotionally unsafe to collaborate digitally with others…(−2)

Table 4: Factor Q sort values sorted by consensus vs. disagreement.

| (id) concept | Factor 1 | Factor 2 | Factor 3 | Factor 4 |
| --- | --- | --- | --- | --- |
| (s5) assertiveness | −1 | 0 | −1 | 0 |
| (s4) interaction management | 0 | 1 | 0 | 1 |
| (s6) altercentrism | 3 | 0 | 1 | 1 |
| (s18) achievement of collective goals | 1 | 2 | 2 | 1 |
| (s8) immediacy | 0 | 1 | 0 | 2 |
| (s19) apprenticeship | 1 | −1 | 2 | 2 |
| (s1) self-disclosure | −1 | −3 | −1 | −2 |
| (s10) environmental control | 1 | 2 | 1 | −1 |
| (s14) cultural differences | −1 | 1 | −2 | 0 |
| (s16) joy and playfulness | 0 | −2 | −1 | 0 |
| (s15) institutional challenges | −2 | −1 | 0 | −1 |
| (s12) technical/operational differences | −3 | 0 | 0 | 0 |
| (s17) exploration and interest | 2 | 1 | −2 | 1 |
| (s11) negative judgment towards other | −1 | 0 | −3 | 0 |
| (s13) emotional barriers | −2 | −2 | 1 | −2 |
| (s2) empathy | 0 | −1 | 0 | 3 |
| (s9) supportiveness | 2 | 3 | −1 | −1 |
| (s7) expressiveness | 0 | −1 | 3 | −1 |
| (s3) social relaxation | 1 | 0 | 0 | −3 |

Three statements differentiating type 2 are S9 (supportiveness: +3), S14 (cultural differences: 1), and S19 (apprenticeship: −1). These three statements demonstrate the significance of cultural differences (+1) that can impede collaboration for members of type 2 and also the low relevance of apprenticeship-related activities in triggering intergenerational collaboration.

*4.1.3. Type 3: Goal-Oriented Younger Collaborators.* Type 3 represents 11 percent of the study variance, with all exemplars being members of the younger generation. Type 3 stresses the design preferences for the expressiveness of the chatbot relative to other statements. For type 3, similar to type 2, an articulate chatbot promotes intergenerational collaboration. Type 3's highest-rated statements are as follows:

(i) s7: (expressiveness) the chatbot should be able to express its feelings either…(+3)

(ii) s18: (achievement of collective goals) I think it is important for me that I can work with others…(+2)

(iii) s19: (apprenticeship) I think it is important that I work together with others…(+2)

Statements are as follows: S7 (expressiveness: +3), S13 (emotional barrier: +1), and S15 (institutional barrier: 0), S17 (exploration), as well as S7 (expressiveness: +3), S13 (emotional barrier: +1), S15 (institutional barrier: 0), and S17 (exploration). Type 3 is also defined by a negative prioritization of the following statements and thus represents cultural openness to collaboration with senior adults despite pursuing different interests:

(i) s11: (negative judgment towards other generation) digital collaboration with other generations…(−3)

(ii) s14: (cultural barriers) cultural differences hinder digital collaboration between…(−2)

(iii) s17: (exploration and interest) I think it is important for me that I can work with others…(−2)

Type 3 showed a significant negative prioritization of cultural barriers and a hostile attitude towards other generations. Members of type 3 also strongly respect senior adults for their apprenticeships (+2).

*4.1.4. Type 4: Tech-Savvy Younger Collaborators.* Type 4 collaborators consisted of three younger people, comprising 10% of the study variance. Type 4 differs from type 3—which also consists exclusively of the younger generation—by classifying two technical (chatbots) interventions as high-ranking statements rather than prioritizing interactive activities. For apprenticeship, the use of chatbots for intergenerational collaboration can be useful for type 4. Type 4 also considers chatbots to facilitate collaboration activities by concentrating on immediate responses, delivering user awareness, and demonstrating empathy. Type 4's top-rank statements are as follows:

  (i) s2: (empathy) chatbot should be able to demonstrate that it understands…(+3)

  (ii) s19: (apprenticeship) I think it is important that I work together with the other generations…(+2)

  (iii) s8: (immediacy) chatbot should be available and open for communication (+2)

The list of differentiated statements shows that type 4 collaborators expect fast responses and social empathy from chatbots. In type 4, the statements are unique according to each member: s2 (empathy: +3), s8 (immediacy: +2), s16 (joy and playfulness: 0), s14 (cultural differences: 0), s10 (environmental control: −1), s13 (emotional barriers: −2), and s3 (social relaxation: −3). Type 4 members are younger collaborators who feel psychologically safe when working with senior adults. Therefore, they do not consider prioritizing chatbots that offer social relaxation. However, type 4 is open to recreational activities moderated by chatbots to support collaboration with senior adults. The following negative statement rankings confirm this:

  (i) s3: (social relaxation) the chatbot should feel comfortable and secure during the in…(−3)

  (ii) s1: (self-disclosure) the chatbot should be able to share personal thoughts or exp…(−2)

  (iii) s13: (emotional barriers) I feel emotionally unsafe to collaborate digitally with others…(−2)

We will now discuss the implications of these results for the intergenerational workplace.

## 5. Discussion

Following the results, both knowledge and practical inputs were discussed for further investigation. As a primary contribution, the results showed four different viewpoints of intergenerational collaborators. Our research complements the chatbot literature [24, 26, 37–39], especially on intergenerational collaboration and the use of chatbot technologies to foster collaboration by incorporating problems and (social activity-based) solutions [40] into each group. The consensus notes that assertive chatbots receive less attention in developing chatbots based on opinion patterns. All identified factors concur that chatbots capable of understanding their own (chatbot) privileges and responsibilities without restricting users' rights are less necessary for intergenerational collaboration. Moreover, there is a consensus in both generations on the reduced relevance of chatbots' assertiveness, showing that integrating the concept of assertiveness into a digital interactive workspace, especially for chatbots, is not fit for an intergenerational environment.

Based on the existing consensus and the different perspectives of the four collaborator types, three recommendations for further research on chatbots and new technologies to facilitate intergenerational collaboration can be made. The results demonstrated that only senior adults are type 1 members. Types 3 and 4 consist of younger adults, and only type 2 members belong to both generations. The "young" perspective comes from type 3 and type 4, while factor 1 comes from an "older" perspective that is very enthusiastic about intergenerational collaboration (e.g., priority for s17: "I think it is important for me to be able to collaborate with others…"), with a preference for a nonjudgmental chatbot (s9: "Chatbot should not judge the user and make us…"). The significant differences between young and older views are an important source for comparative analysis, as the cumulative percentage of types 3 and 4 as a combination also accounts for up to 21% of the preference pattern (type 3 with 11% + type 4 with 10%), almost reaching the same percentage as 23% for type 1. Therefore, our research's first proposition focuses on understanding the composition of each category of collaborators:

**Proposition 1.** *Young adults' preference for chatbots in the collaboration space is more diverse than senior adults' preferences. Chatbot system designers should prioritize the design preferences of senior adults over the preferences of younger adults to enable most senior adults' participation.*

Proposition 1 contrasts with previous studies, suggesting that senior adults are heterogeneous. Consequently, their inclusion requires a multifaceted approach [1, 10, 58]. The senior adult group tends to have a more unified perception of the study context (in type 1) than younger adults. One reason for this could be senior adults' different experiences in interacting with other generations [2, 51, 59]. Regardless of the mediation technology, senior adults with rich experiences have a more positive intergenerational cooperation point of view.

Therefore, we argue that Proposition 1 could impact the understanding of social robots' use for intergenerational collaboration. First, the organization or system designer should find common ground for chatbots (or any other type of social robot) in a collaborative workspace between the two generations. Secondly, organizations should prioritize senior adults when designing intergenerational collaboration support systems. Given the different types of limitations for senior adults, our findings reinforce the concept of designing social inclusion [52, 53]. Suppose that people with limitations and special needs can use the proposed technology. In that case, those without limitations are also capable of using it. Hence, the design is expected to cover a broader and more diverse range of users by applying Proposition 1 [52, 53].

In a similar vein, we examine the order of the statements' position in Table 4: the top two activity category statements are statements 18 and 19, indicating that both activities are closer to consensus than the other activity categories. Both statements have positive ratings overall among all collaborator types. Therefore, based on the results, the following proposition is derived:

**Proposition 2.** *Training for apprenticeships and achieving collective goals are two critical collaborative activities that chatbots can support to foster joint collaborative innovation between different adult generations.*

Proposition 2 consists of exchanging knowledge to promote intergenerational collaboration among different collaborators. In comparison, mutual values and playful activities may also encourage intergenerational collaboration [14, 17]. This research helps explain that the focus should be on learning and capability development in chatbot-mediated communication within intergenerational settings. Thus, Proposition 2 affirms earlier research promoting capacity building in intergenerational partnerships [2, 26, 51, 59].

Now, to determine the pattern of the barrier statements. All barrier variables were viewed as negative neutral, excluding emotional barriers (+1 in type 3). The pattern indicates that chatbots could solve most barriers to intergenerational collaboration. Three categories of collaborators (type 1, type 2, and type 4) gave unfavorable scores to emotional barriers compared to other barriers. Hence, we postulate the following:

**Proposition 3.** *Chatbots are well suited to intergenerational collaboration, but chatbots do not (are not required to) facilitate emotional bonding between collaborators.*

Proposition 3 contains two knowledge implications. First, our study results support previous studies by validating the fact that chatbots can be used to strengthen workplace collaboration [25, 38] and raise chatbot awareness for intergenerational use. Second, this research limits the value of chatbots' emotional attachment across age-based generations by considering the context of the activity. In our research, chatbots' emotional attachment is less significant, as collaboration focuses on work-oriented activities.

In terms of practical contributions, the proposed types of collaborators can also be defined as "personas" [19, 40, 41]. Personas describe the system designer's imaginary users, who assist the system designer's goal-oriented design [41]. Cooper et al., [41] classify personas into primary, secondary, and complementary personas. The four types of collaborators start with type 2 as the primary personas since type 2 is associated with both generations. Type 2, which consists of both generations, also supports the collaborating personas [40]. Type 2 does not only describe a specific group of generations. Type 1 – as a secondary persona representing the senior generations and is more open towards intergenerational collaboration. Types 3 and 4 as complementary personas indicate that further support is required to enhance intergenerational collaboration. Figure 3 is an illustration of the personas based on the type of collaborator.

Furthermore, the study results can also help chatbot developers decide which design features associated with a particular chatbot concept should be developed first within the intergenerational context by ranking each collaborator's statements. Due to a lack of resources, prioritizing design features is essential to enabling positive digital experiences [11, 56, 82], 2015 for both intergenerational innovation and startups. By using a lean approach [49], system designers can focus on features that are essential for user-design requirements—for example, developing a feature that supports chatbots' (s4) interaction management and (s6) altercentrism, as both statements tend to be consensus statements and have a positive rank among different collaborators. Chatbot designers can choose a statement that selects for type 2 by focusing on a concept that has reached a higher rank (s9—supportiveness: +3) and extending functions to positively ranked concepts (environment control: +2, immediacy: +1). Figure 4 shows the proposed chatbots (prototype) based on the previous analysis of the primary persona (type 2).

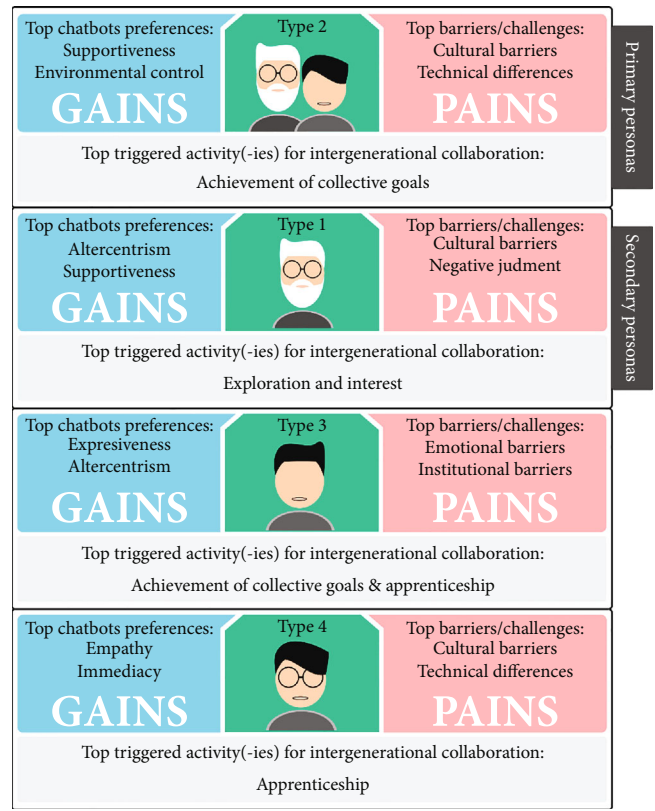

FIGURE 3: Personas of intergenerational collaborators for chatbots.

Moreover, as a practical contribution, we demonstrated how the Q methodology represents an alternative system-design tool [53]. This methodology can be used for user experience designers [83] to find pattern similarities in different sociotechnical viewpoints used in social inclusion design [52, 53, 55, 84], and also, this study demonstrates the potential of the Q method to identify preferred interaction patterns that can be used to improve personalization and customization of chatbots [22]. Based on our findings, we introduce the idea of using chatbots focused on subjective preferences independently of age and social differences. Therefore, we propose the following interventions in the Q

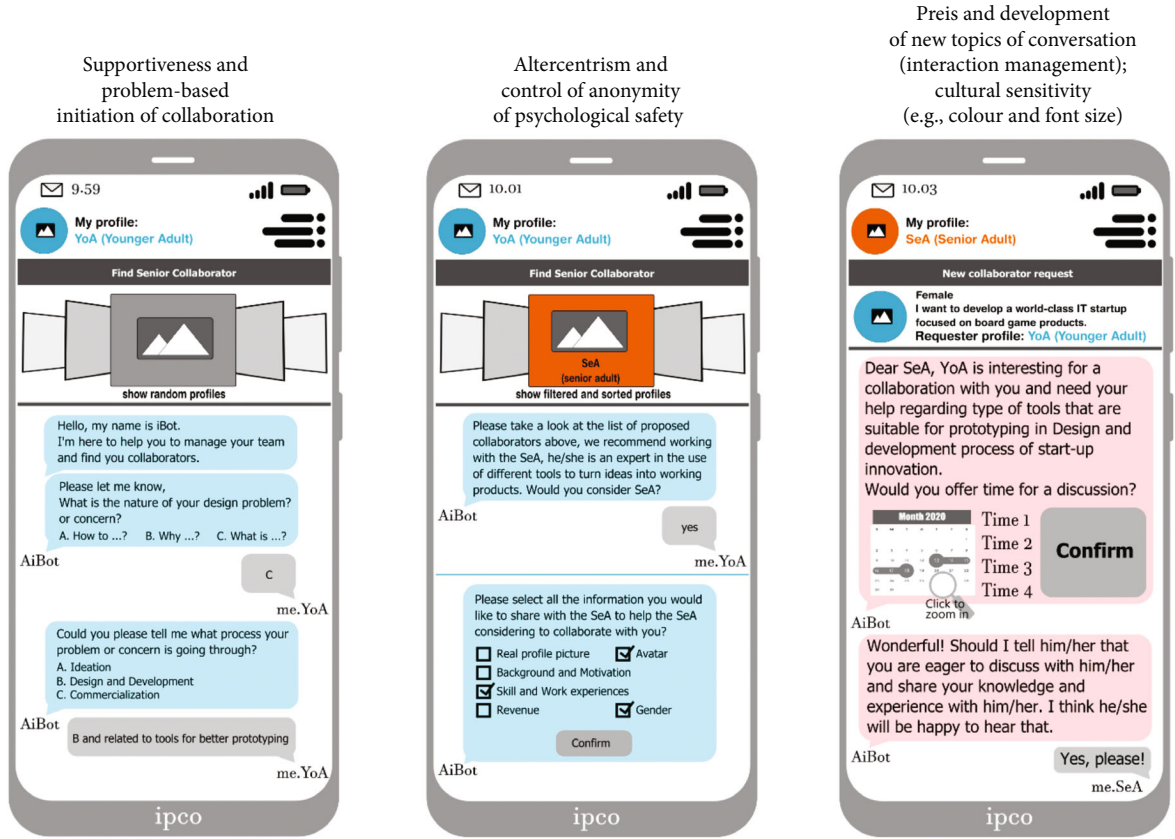

Figure 4: Proposed chatbots prototyped for type 2.

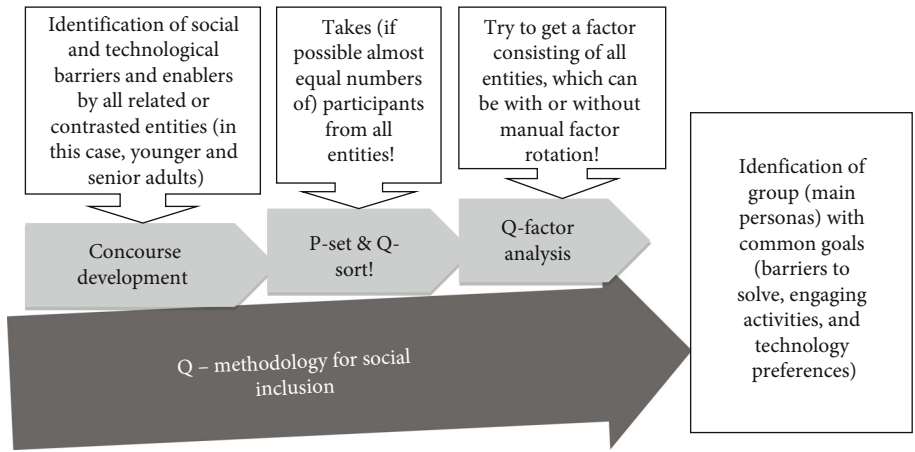

Figure 5: The adoption of the Q methodology to design systems for social inclusion.

methodology process to support social inclusion in technology development (see Figure 5).

Our study demonstrates how the Q methodology can quantify subjectivity in the early phases of product and system development [54, 84]. We also support previous studies recommending adopting the Q methodology to analyze design requirements and system affordances [44, 45, 74, 75, 84, 85], especially in the design science research process. Accordingly, we also encourage researchers to utilize the Q methodology to explore human perspectives for technology design.

## 6. Conclusions, Limitations, and Recommendations

The diversity of the workforce is one of the main drivers of innovation. Given the upcoming trend—triggered by AI advancements—of using chatbots in the digital work environment, chatbots as social robots could be an attractive alternative technology to moderate collaborative innovation activity. In summary, this study has shed some light on the potential of chatbots and different types of collaborators in producing intergenerational innovation. This study contributes theoretically and practically to the literature by systematically identifying the subjectivity pattern associated with chatbot design using the Q methodology.

Nevertheless, some drawbacks of the analysis can be changed and considered for further studies. The essence of the Q methodology allows us to perform experiments in a particular setting without needing a broad sample of subjects to generalize the findings. Consequently, the observation is constrained and cannot be extrapolated to other contexts of technology collaboration. Furthermore, the Q methodology's potential was not fully optimized since only text-based Q sets were used. Alternatively, wireframes, images, or videos can build visual Q sets to verify specifications and design interactions.

We developed a literature-based concourse exclusively for the Q set. Future research using the Q method to design chatbots or other technological interventions may develop Q sets from direct user surveys and observations, social media analysis, or expert interviews. Accordingly, the Q set may be more complicated. However, it will be rich in practical design features and hence find more patterns. For example, game elements support the playful environment and convince the system design to promote collaborative behavior across both generations.

## Data Availability

Data for this study are available from the author upon request.

## Conflicts of Interest

The authors declare that they have no conflicts of interest.

## Acknowledgments

Irawan Nurhas's work is supported by the Ministry of Culture and Science of the State of North Rhine-Westphalia and open-access funding enabled and organized by the Institute of Positive Computing at Hochschule Ruhr West University of Applied Sciences.